----------
X-Sun-Data-Type: default
X-Sun-Data-Description: default
X-Sun-Data-Name: moshe1
X-Sun-Charset: us-ascii
X-Sun-Content-Lines: 618

\magnification=1200
\settabs 18 \columns

\baselineskip=12 pt
\topinsert \vskip 1.25 in
\endinsert

\def\sqr#1#2{{\vcenter{\vbox{\hrule height.#2pt
 \hbox{\vrule width.#2pt height#1pt \kern#1pt
 \vrule width.#2pt} \hrule height.#2pt}}}}

\def\operp{\hbox{${\kern+.25em{\bigcirc}
\kern-.85em\bot\kern+.85em\kern-.25em}$}}

\def\lsim{\;\raise0.3ex\hbox{$<$\kern-0.75em\raise-1.1ex\hbox{$\sim$}}\;}
\def\gsim{\;\raise0.3ex\hbox{$>$\kern-0.75em\raise-1.1ex\hbox{$\sim$}}\;}
\def\no{\noindent}

\def\ce{\centerline}
\def\ve{\vfill\eject}
\def\rdots{\mathinner{\mkern1mu\raise1pt\vbox{\kern7pt\hbox{.}}\mkern2mu
 \raise4pt\hbox{.}\mkern2mu\raise7pt\hbox{.}\mkern1mu}}

\def\e e{$e^+ e^-$ }



\rightline{UCLA/99/TEP/20}
\rightline{May 1999}
\vskip1.0cm

\ce{{\bf OBSERVABLE PROPERTIES OF $q$-DEFORMED PHYSICAL SYSTEMS}}
\vskip.5cm
\ce{\it R. J. Finkelstein}
\vskip.3cm
\ce{Department of Physics and Astronomy}
\ce{University of California, Los Angeles, CA  90095-1547}
\vskip1.0cm

\ce{\it To Mosh\'e Flato}
\vskip1.0cm

\no {\bf Abstract.}  Examples are given of $q$-deformed systems
that may be interpreted by the standard rules of quantum theory in terms
of new degrees of freedom and supplementary quantum numbers.
\ve

\line{{\bf 1. Introduction.} \hfil}
\vskip.3cm

In standard quantum mechanical theories the $S$-matrix encodes the physically
observable properties: transition probabilities between scattering-states and energy levels of bound states.  In this note we are
similarly concerned with energy levels and transition probabilities of
physical systems which have been $q$-deformed.  The examples we consider
are speculative deformations of real physical systems which have been
extensively discussed in diverse ways.  Because of the consequent 
arbitrariness and ambiguities in these different approaches it is perhaps
useful to attempt a more uniform interpretation of a few typical cases.
\vskip.5cm

\line{{\bf 2. The $q$-Oscillator.$^1$} \hfil}
\vskip.3cm

In the Fock representation the $q$-oscillator is defined by the Hamiltonian
$$
H = {1\over 2} (a\bar a + \bar aa) \hbar\omega \eqno(2.1)
$$
\no where $\bar a$ and $a$ satisfy
$$
(a,\bar a)_q = a\bar a-q\bar aa = 1~. \eqno(2.2)
$$
\no The time dependence of these operators is given by the $q$-Dirac rule:
$$
i\hbar \dot a = (a,H)_q \to a\sim e^{-i\omega t} \eqno(2.3)
$$
Operating on eigenstates of $H$ the $a$ and $\bar a$ satisfy
$$
\eqalignno{a|n\rangle &= \langle n\rangle^{1/2} |n-1\rangle & (2.4a) \cr
\bar a|n\rangle &= \langle n+1\rangle^{1/2} |n+1\rangle & (2.4b) \cr}
$$
\no and
$$
H|n\rangle = {1\over 2}\bigl[\langle n\rangle +
\langle n+1\rangle\bigr] \hbar\omega \eqno(2.5)
$$
\no where $\langle n\rangle$ is the basic number
$$
\langle n\rangle = {q^n-1\over q-1}~. \eqno(2.6)
$$
\no Denote the ground state by $|0\rangle$.  Then the $n^{\rm th}$ state is
$$
|n\rangle \sim \bar a^n|0\rangle~. \eqno(2.7)
$$
Set
$$A = \left(\matrix{a\cr \bar a\cr}\right)~. \eqno(2.8)
$$
\no Then (2.2) may be rewritten as
$$
A^t \epsilon A = q^{-1/2} \eqno(2.9)
$$
\no where
$$
\epsilon = \left(\matrix{0 & q^{-1/2} \cr
-q^{1/2} & 0 \cr} \right)~. \eqno(2.10)
$$
\no One may transform to other representations by the following canonical
transformation, $T$, belonging to $SU_q(2)$
$$
A = TX~, \quad T\epsilon SU_q(2) \eqno(2.11)
$$
\no where $T$ satisfies
$$
T^t \epsilon T = T\epsilon T^t = \epsilon~. \eqno(2.12)
$$
\no Then
$$
X^t\epsilon X = q^{-1/2} \eqno(2.13)
$$
\no and if one sets
$$
X = \left(\matrix{{i\over \hbar} p \cr x \cr} \right) \eqno(2.14)
$$
\no then
$$
qxp-px = i\hbar~. \eqno(2.15)
$$
These relations may be satisfied in configuration space by the pair
$(-i\hbar D^q_x,x)$ or in momentum space by the pair $(p,-iq_1D_p^{q_1})$
where the $D$ are the following difference operators:
$$
\eqalignno{D_x^q\psi(x) &= {\psi(qx)-\psi(x)\over qx-x} & (2.16a) \cr
D^{q_1}_p \varphi(p) &= {\varphi(q_1p)-\varphi(p)\over q_1p-p} & (2.16b) \cr}
$$
\no and
$$
q_1 = q^{-1}~. \eqno(2.17)
$$
Set
$$
T = \left(\matrix{\alpha & \beta \cr -q_1\bar\beta & \bar\alpha}\right)~.
\eqno(2.18)
$$
\no Then (2.12) implies
$$
\eqalign{\alpha\beta = q\beta\alpha ~~ {\rm (a)} \qquad 
&\alpha\bar\alpha + \beta\bar\beta = 1 ~~~~~
{\rm (d)} \cr
\alpha\bar\beta = q\bar\beta\alpha ~~ {\rm (b)} \qquad
&\bar\alpha\alpha + q_1^2\bar\beta\beta = 1 ~~ {\rm (e)} \cr
\beta\bar\beta = \bar\beta\beta ~~~~ {\rm (c)} \qquad
& \hfil \cr} \eqno(2.19)
$$
\no By (2.11)
$$
\eqalign{a &= \alpha\biggl({i\over \hbar} p\biggr) + \beta x \cr
\bar a &= -q_1\bar\beta\biggl({i\over\hbar} p\biggr) + \bar\alpha x~. \cr}
\eqno(2.20)
$$
\no By (2.7) and (2.20) the eigenstates may be expressed in either the
$x$ and $p$ representation and are the products of $q$-Gaussian and
$q$-Hermite polynomials.$^1$  These expressions will depend upon basic
numbers and some version of basic Hermite functions in either $x$ or $p$
and will agree with the usual expressions in the $q=1$ limit.  The $x$ and $p$ amplitudes are related by
$q$-Fourier transforms:
$$
\eqalign{\psi_n(x) &= \int^\infty_{-\infty}
{\cal{E}}_q(ipx)\varphi_n(\rho)d_qp \cr
\varphi_n(x) &= \int^\infty_{-\infty}
{\cal{E}}_{q_1}(-iq\rho x) \psi_n(x) d_qx \cr}
$$
\no where ${\cal{E}}_q(ipx)$ is an eigenstate of $p$:
$$
{\cal{E}}_q(ipx) = \sum
{(ipx)^n\over\langle n\rangle!}
$$
\no and the $q$-integrals are Jackson integrals (inverse
of $q$-differentiation).

However,   
all of these functions lie in the subalgebra generated by $\alpha$ and
$\beta$.  That basic numbers and basic hyper-geometric functions replace
the usual expressions does not concern us.  On the other hand, that the wave
functions lie in an algebra and are not numerically valued requires
attention.

If the operator associated with a transition between two states, $m$ and
$n$, is $A$, then the usual transition amplitude is $\langle n|A|m\rangle$
and the corresponding transition probability is $|\langle n|A|m\rangle|^2$;
but if the states as well as the operator lie in the algebra,
$|\langle n|A|m\rangle|^2$ does not have its usual interpretation since it
is not numerically valued.  There are two natural procedures for dealing
with the problem.

The first way to attach a numerical measure to $|\langle n|A|m\rangle|^2$
is to form the Woronowicz measure as follows:
$$
h|\langle n|A|m\rangle|^2 \eqno(2.21)
$$
\no where $h$ stands for the Woronowicz integral over the algebra.  It may
be evaluated term by term with the aid of the following result:
$$
h[\alpha^s\beta^n\bar\beta^m] = \delta^{so}\delta^{mn}q^n/[m+1]_q
\eqno(2.22)
$$
\no where
$$
[m] = {q^m-q^{-m}\over q-q^{-1}}~. \eqno(2.23)
$$
\no On top of this integration there is the usual $x$ or $p$ integration.

In an alternative procedure, which is also natural since it is also based
on the $q$-algebra, we construct a Hilbert space as follows:

Because $\beta$ and $\bar\beta$ commute they have common eigenstates.  Let
$|0\rangle$, the groundstate of this Hilbert space, be a common eigenstate
of $\beta$ and $\bar\beta$.  Then
$$
\eqalign{\beta |0\rangle &= b|0\rangle \cr
\bar\beta |0\rangle &= \bar b|0\rangle \cr
\alpha |0\rangle &= 0~. \cr} \eqno(2.24)
$$
\no By (2.19e) it follows that $|b|^2=q^2$.

To see that $\bar\alpha$ and $\alpha$ are consistent creation and destruction
equations define
$$
|n\rangle \sim \bar\alpha^n|0\rangle~. \eqno(2.25)
$$
\no By iteration of (2.19a)
$$
\beta|n\rangle = (q^nb)|n\rangle~.  \eqno(2.26)
$$
\no Then $|n\rangle$ is an eigenstate of $\beta$ and $\bar\beta$ and
$$
\eqalignno{\bar\alpha|n\rangle &= \lambda_n|n+1\rangle & (2.27) \cr
\alpha|n\rangle &= \mu_n|n-1\rangle & (2.28) \cr}
$$
\no where
$$
\eqalignno{\lambda_n &= (1-|b|^2q^{2n})^{1/2} & (2.29) \cr
\mu_n &= \lambda_{n-1}~. & (2.30) \cr}
$$

If $\alpha$ and $\bar\alpha$ are adjoint and the Hilbert space has a positive
definite scalar product, then $\langle n|\alpha\bar\alpha|n\rangle$ and
$\langle n|\bar\alpha\alpha|n\rangle$ must remain positive, and therefore
$q^2\leq 1$ by Eqs. (2.19d) and (2.19e). 
Continuing this procedure one now replaces
$\langle N|A|M\rangle$ by $\langle nN|A|mM\rangle$ where $n$ and $m$ refer to
states of the associated Hilbert space.  Now a state of the $q$-oscillator
carries two quantum numbers, $n$ and $N$, and the matrix elements are
numerically valued.  Attaching a new quantum number, $n$, is analogous
to the procedure in Pauli spin theory where one multiplies an atomic wave
function by an independent spin function.  The $q$-oscillator interpreted
in this way has internal degrees of freedom.
\vskip.5cm

\line{{\bf 3. The $q$-Coulomb Problem (The $q$-Hydrogen Atom).$^2$} \hfil}
\vskip.3cm

The Schr\"odinger differential equation in configuration space becomes an
integral equation in momentum space.  To emphasize the $O(3)$ symmetry
of the Coulomb problem one may map momentum space onto the group space of
$O(3)$ according to
$$
{p_o-i\vec p~\vec \sigma\over p_o + i\vec p~\vec\sigma} =
e^{{1\over 2}i\vec\sigma~\vec w} \eqno(3.1)
$$
\no where $p_o$ and $\vec p$ refer to the energy and momentum while
$\vec w$ fixes the magnitude and axis of rotation.  Since the problem is
non-relativistic the relativistic bound state energy $E$ is
$$
E = -{p_o^2\over 2m}~. \eqno(3.2)
$$
\no It is then possible to cast this integral equation into the following
form:
$$
\int K(\vec p,\vec p^\prime) \Phi(\vec p^\prime) d\tau(\vec p^\prime) =
C p_o\Phi(\vec p) \eqno(3.3)
$$
\no where
$$
C = {\hbar\over me^2}
$$
\no and
$$
K(\vec p,\vec p~^\prime) = \sum_{jm\mu} 
\bar D^j_{m\mu}(\vec p)
D^j_{m\mu}(\vec p~^\prime)~. \eqno (3.4)
$$
\no Here $K$ is a rescaled Fourier transform of the potential and $\Phi$
is the rescaled wave function.  The $D^j_{m\mu}(\vec p^\prime)$ are
Wigner functions (matrix elements of irreducible representations of the
rotation group).  These functions satisfy the orthogonality relations
$$
\int \bar D^j_{m\mu}(\vec p) D^{j^\prime}_{m^\prime\mu^\prime}(\vec p)
d\tau(\vec p) = \delta^{jj^\prime}\delta_{mm^\prime}
\delta_{\mu\mu^\prime}/2j+1~. \eqno(3.5)
$$
\no By (3.3), (3.4), and (3.5) one sees that the eigenfunctions of (3.3) are
$$
\Phi^j_{m\mu}(\vec p) = D^j_{m\mu}(\vec p) \eqno(3.6)
$$
\no and the eigenvalues are
$$
Cp_o = {1\over 2j+1} \eqno(3.7)
$$
\no and by (3.2) one finds the Balmer formula for the energy
$$
E = -{1\over 2m}{1\over e^2}{1\over (2j+1)^2} =
-{1\over 2} mc^2\biggl({e^2\over \hbar c}\biggr)^2 {1\over N^2} \eqno(3.8)
$$
\no where the principal quantum number is
$$
N = 2j+1~. \eqno(3.9)
$$
\no The other indices $m$ and $\mu$ labelling $\Phi^j_{m\mu}$ refer to the
$z$-components of the Lenz and angular momentum vectors.  These are the
conserved integrals of the motion.  The $O(4)$ algebra of this system is
reflected in the $O(3)\times O(3)$ symmetry of the left and right parameter
groups.

The amplitude in momentum space $\varphi(\vec p)$ is related to
$\Phi(\vec p)$ by a rescaling as follows:
$$
\varphi^j_{mm^\prime}(\vec p) = G^2(p) D^j_{mm^\prime}
(\vec p)~. \eqno(3.10)
$$
\no Here $G(p)$ is the ``propagator":
$$
G(p) = {p_o\over p^2 + p_o^2}~. \eqno(3.11)
$$

We shall now define the $q$-Coulomb problem by deforming the integral
equation (3.3).  To formulate this new equation it is necessary to define
integration over the ``$q$-group space".  This may be done by again making
use of the Haar measure introduced by Woronowicz who established orthogonality
relations that may be expressed as follows:
$$
h(\bar D^j_{m\mu} D^{j^\prime}_{m^\prime\mu^\prime}) =
\delta^{jj^\prime}\delta_{mm^\prime}\delta_{\mu\mu^\prime}
q^{2m}/[2j+1]_q \eqno(3.12)
$$
\no where $h$ is the Haar measure and $D$ a unitary representation.  In our
notation the preceding equation is 
$$
\int_W \bar D^j_{m\mu}(\alpha|q) D^{j^\prime}_{m^\prime\mu^\prime}
(\alpha|q) d\tau(\alpha) = \delta^{jj^\prime}\delta_{mm^\prime}
\delta_{\mu\mu^\prime}q^{2m}/[2j+1]_q \eqno(3.13)
$$
\no where $f(\alpha)$ lies in the algebra generated by $(\alpha,\bar\alpha,
\beta,\bar\beta)$, $\int_W$ means a Woronowicz integral, and
$D^j_{m\mu}(\alpha|q)$ is a matrix element of an irreducible representation
of the ``$q$-group".  The deformed integral equation corresponding to (3.3) is
$$
\int_W K_q(\alpha,\alpha^\prime)\Phi(\alpha^\prime,q)
d\tau(\alpha^\prime) = Cp_o\Phi(\alpha|q) \eqno(3.14)
$$
\no where the kernel is
$$
K_q(\alpha,\alpha^\prime) = \sum D^j_{m\mu}(\alpha|q)
\bar D^j_{m\mu}(\alpha^\prime|q)~. \eqno(3.15)
$$
\no Utilizing the orthogonality relations as before one finds that
$$
\Phi(\alpha|q) = D^j_{m\mu}(\alpha|q) \eqno(3.16)
$$
\no and
$$
E(N,\mu) = -{1\over 2} mc^2\biggl({e^2\over \hbar c}\biggr)^2
{q^{4\mu}\over [N]^2} \eqno(3.17)
$$
\no where $N(=2j+1)$ is again the principal quantum number.  Here
$$
[N] = {q^N-q^{-N}\over q-q^{-1}}~,
$$
\no inherited from (3.12).

In the $q=1$ limit one recovers the usual results.  If $q\not=1$, the
expression for the energy is modified, similar to the way it was for
the oscillator, by the substitution of the basic number $[N]$ for $N$.

The factor $q^{4\mu}$ reveals that the Coulomb degeneracy is lifted and
that the $q$-system no longer has spherical symmetry.

Since the wavefunctions $D^j_{m\mu}(\alpha|q)$ lie in the algebra, one has
the same problem in computing transition probabilities that one has for
the oscillator or for any other problem obtained by $q$-deformation; and
one can handle the problem in the ways suggested for the oscillator.  
Probably the simpler procedure is to evaluate these probabilities as 
squares of matrix
elements between states of the Hilbert space defined by the algebra.  In
doing this we are explicitly recognizing that the $q$-system has new
degrees of freedom not belonging to the $q=1$ system but we are not discarding
the usual quantum mechanical rules of interpretation.

This Hilbert space is intrinsic to the formalism and in fact is very similar
to the usual Fock space.  The suggested procedure is analogous to the
evaluation of operator fields in Fock space, i.e., the $q$-deformed problem
is more like a quantum field problem than a one-particle quantum mechanical
problem.

The $q=1$ oscillator as well as the $q=1$ hydrogen atom are themselves
idealizations of real physical systems.  Whether these new $q\not=1$ systems
are also useful idealizations of more complex physical systems is not known.
\vskip.5cm

\line{{\bf 4. $q$-Field Theory.$^3$} \hfil}
\vskip.3cm

An arbitrary field may be expanded as follows:
$$
\psi_\mu(x) = \sum_\rho [a(\rho) f_\mu(\rho,x) + \bar b(\rho) g_\mu(\rho,x)] \eqno(4.1)
$$
\no where $\mu$ is a generic tensor index and
$$
\eqalignno{&\sum_\rho = \sum_r \int d\vec p 
\qquad ~~~\rho = (\vec p,r) & (4.2a) \cr
& f_\mu(\rho,x) \sim {u_\mu(\vec p,r)\over (2p_o)^{1/2}} e^{-ipx} & (4.2b) \cr
& g_\mu(\rho,x) \sim {v_\mu(\vec p,r)\over (2p_o)^{1/2}} e^{ipx}~. 
& (4.2c) \cr}
$$
\no Here $a(\bar a)$ and $b(\bar b)$ are absorption (emission) operators
of particles and antiparticles respectively.  The $\rho$ sum is an integration
over momentum and a sum over spin states ($r$).  The particle and antiparticle
parts of the sum are related by complex conjugation of the exponentials
and by charge conjugation of the spin dependent functions.

Let us now impose the $q$-commutators
$$
\eqalignno{(a(p),~\bar a(p^\prime))_q &= \delta(p,p^\prime) & (4.3a) \cr
(b(p),~\bar b(p^\prime))_q &= \delta(p,p^\prime) & (4.3b) \cr
(a(p),~b(p^\prime))_q &= 0 & (4.3c) \cr
(\bar b(p),~\bar a(p^\prime))_q &= 0 & (4.3d) \cr}
$$
\no In standard theory $q=1~(-1)$ describes the (boson) fermion fields.

One may also define the $q$ time-ordered product
$$
\eqalign{T_q(\psi(x),\psi(x^\prime)) &= 
\psi(x) \psi(x^\prime) \qquad ~t>t^\prime \cr
&= q\psi(x^\prime)\psi(x) \qquad t<t^\prime \cr} \eqno(4.4)
$$
\no again correct for both bosonic and fermionic fields.  Finally introduce
the $q$-modified $S$ matrix:
$$
S^{(q)} = T_q\biggl(\exp\bigl(i\int L(x) d^4x\bigr)\biggr)~. \eqno(4.5)
$$
\no In the standard way, we take the transition probability between an
incoming state $A$ and an outgoing state $B$ to be 
$|\langle B|S^{(q)}|A\rangle|^2$ where $A^+$ and $B^+$ 
are products of creation operators that represent different 
collections of particles.  Here
$$
\eqalignno{|A\rangle &= A^+|0\rangle & (4.6a) \cr
\langle B| &= \langle 0|B & (4.6b) \cr
\langle B|S|A\rangle &= \langle 0|BSA^+|0\rangle & (4.6c) \cr}
$$

Again following the usual procedures $\langle B|S^{(q)}|A\rangle$ is
expanded and evaluated with the aid of Wick's theorem for normal products.
In putting a string of absorption and emission operators in normal form
one picks up a power of $q$ instead of a power of -1.  Continuing this
procedure one arrives at a set of Feynman rules differing from the
standard rules by the substitution of $q$-dependent internal propagator
for the usual internal propagators.

One finds for the vector fields
$$
D_{\mu\lambda}(x) = \biggl(g_{\mu\lambda}-{\partial_\mu\partial_\lambda\over
m^2}\biggr)
\biggl({1\over 2\pi}\biggr)^4 {1+q\over 2} \int e^{-ikx}
\biggl({1\over k^2-m^2} + {1-q\over 1+q}{k_o\over\omega}\biggr)
d^4k \eqno(4.7)
$$
\no and for spinor fields
$$
S_{\alpha\beta}(x) = (\partial \!\!\!/ + m)_{\alpha\beta}
\biggl({1\over 2\pi}\biggr)^4\biggl({1-q\over 2}\biggr)
\int e^{-ikx}
\biggl[{1\over k^2-m^2} + {1+q\over 1-q}{k_o\over\omega}\biggr]
d^4k \eqno(4.8)
$$
\no where
$$
\omega = (\vec k^2+m^2)^{1/2}~. \eqno(4.9)
$$
\no For the standard vector (spinor) field $q=1(-1)$ these propagators reduce
to their standard form.

One may test these expressions by computing, for example, electron-electron
scattering or electron-positron annihilation.  One obtains the standard
(M\"oller) expression for electron-electron scattering and the standard
(Bhabha) expression for electron-positron annihilation but both expressions
are multiplied by frame dependent factors.

In this way one sees that $q$-quantization imposed in the minimal way we
have described breaks the Lorentz symmetry.  A somewhat similar result was 
obtained in discussing the $q$-Coulomb problem where the $q$-quantization
(there differently implemented) breaks the rotational symmetry.  The propagators
(4.7) and (4.8) also illustrate the Pauli result that the Lorentz group
requires $q=1$ for bosons and $q=-1$ for fermions.

The preceding result suggests that the argument be reversed by postulating
a $q$-symmetry group at the outset.
One may therefore introduce the $q$-Lorentz group by defining its spin
representation $L_q$ as follows:
$$
\epsilon_q~{\rm det}_q L_q = L^t_q\epsilon_qL_q = L_q\epsilon_qL_q^t~, \quad
{\rm det}_q L_q=1~. \eqno(4.10)
$$
\no Here $\epsilon_q$ is the $q$-deformed Levi-Civita matrix as before and
Eq. (4.10) is also the definition of the $q$-determinant of $L_q$.
If $q=1$, Eq. (4.10) exactly defines the spin representation of the
Lorentz group.

To see how the postulate of $q$-Lorentz invariance 
might work out, one may
focus on the correlation functions (vacuum expectation values of products
of interacting fields) since these determine the structure of the theory.
Without going into details, the essential new element now introduced by the
$q$-symmetry is that the interacting fields lie in a $q$-algebra and, just
as for the simple systems already discussed, there are new degrees of freedom
that give rise to new quantum numbers determined by the Hilbert space associated
with the algebra.

In a $q$-Yang-Mills theory the connection field would lie in a $q$-algebra
rather than a Lie algebra.  The associated Fock space would then be the
direct product of the standard Fock space generated by $\bar a$ and the
$q$-Fock space generated by $\bar\alpha$.  If Yang-Mills
theories describe point particles, then $q$-Yang-Mills
theories describe particles with internal degrees
of freedom.

I should like to thank Professors Fronsdal and Varadarajan for comments.
\vskip.5cm

\line{{\bf References.} \hfil}
\vskip.3cm

\item{1.} R. Finkelstein and E. Marcus, J. Math. Phys. {\bf 36}, 2652
(1995).
\item{2.} R. Finkelstein, J. Math. Phys. {\bf 37}, 2628 (1996).
\item{3.} R. Finkelstein, J. Math. Phys. {\bf 37}, 983 (1996).

\end
\bye